\documentclass[aps,prd,reprint,superscriptaddress]{revtex4-1}
\usepackage{amsmath,amssymb}
\usepackage[colorlinks=true,allcolors=blue]{hyperref}
\usepackage{dcolumn}
\usepackage{units}
\usepackage{graphicx}
\usepackage{float}
\usepackage{enumitem}
\usepackage{diagbox}
\usepackage{overpic}
\usepackage{wasysym}
\usepackage{comment}
\usepackage{color}

\allowdisplaybreaks[1]

\newcommand{\be}{\begin{equation}}
\newcommand{\ee}{\end{equation}}
\newcommand{\tref}[1] {Table~\ref{#1}}
\newcommand{\eref}[1] {Eq.~(\ref{#1})}
\newcommand{\sref}[1] {Sec.~\ref{#1}}

\newcommand{\ve}[1]{\mathbf{#1}}

\def\eqspa{\ \ , \ \ }

\newcommand{\rrr}[1]{\mathbf{{r}}_{#1}}

\newcommand{\vvv}[1]{\mathbf{{v}}_{#1}}

\newcommand{\PP}{{\mathbf{P}}}

\newcommand{\zero}{\mathbf{0}}
\newcommand{\dertot}[2]{\frac{d{#1}}{d{#2}}}
\newcommand{\dersec}[2]{\frac{d^2{#1}}{d{#2}^2}}
\newcommand{\derparz}[2]{\frac{\partial {#1}}{\partial {#2}}}

\begin{document}
\title{Constraining the Nordtvedt parameter with the BepiColombo \\Radioscience experiment}
\author{Fabrizio De Marchi}
\email[Corresponding author: ]{fabrizio.demarchi@uniroma1.it}
\affiliation{Department of Mechanical and Aerospace Engineering, Sapienza University of Rome,   Via Eudossiana, 18, 00184 Rome, Italy}

\author{Giacomo Tommei}
\affiliation{Department of Mathematics, University of Pisa, Largo Bruno Pontecorvo 6, 56127 Pisa, Italy}

\author{Andrea Milani}
\affiliation{Department of Mathematics, University of Pisa, Largo Bruno Pontecorvo 6, 56127 Pisa, Italy}

\author{Giulia Schettino}
\affiliation{Department of Mathematics, University of Pisa, Largo Bruno Pontecorvo 6, 56127 Pisa, Italy}

\begin{abstract}
BepiColombo is a joint ESA/JAXA mission to Mercury with challenging objectives regarding geophysics, geodesy and fundamental physics. 
The Mercury Orbiter Radioscience Experiment (MORE) is one of the on-board experiments, including three different but linked 
experiments: gravimetry, rotation and relativity. The aim of the relativity experiment is the measurement of the 
post-Newtonian parameters. Thanks to accurate tracking between Earth and spacecraft,  the results are expected to be very precise. 
However, the outcomes of the experiment strictly depends on our "knowledge" about solar system:  ephemerides, number of bodies 
(planets, satellites and asteroids) and their masses. In this paper we describe a semi-analytic model used to perform a covariance analysis
 to quantify the effects, on the relativity experiment, due to the uncertainties of solar system bodies parameters. In particular, 
 our attention is focused on the Nordtvedt parameter $\eta$ used to parametrize the strong equivalence principle violation. \\
 After our analysis we estimated $\sigma[\eta]\lessapprox 4.5\times 10^{-5}$ which is about 1~order of magnitude larger than the "ideal" case 
 where masses of  planets and asteroids have no errors. The current value, obtained from ground based experiments and lunar laser ranging 
 measurements, is $\sigma[\eta]\approx 4.4\times 10^{-4}$.\\
  Therefore, we conclude that, even in presence of uncertainties on solar system parameters, the measurement of $\eta$ by MORE can improve
   the current precision of about 1~order of magnitude.
 \end{abstract}


\maketitle


\section{Introduction}\label{intro}

BepiColombo (BC) is a joint ESA/JAXA mission aimed at the exploration
of Mercury. The mission is composed by two spacecrafts, the ESA
Mercury Planetary Orbiter (MPO) and the JAXA Mercury Magnetospheric
Orbiter (MMO), to be put in orbit around Mercury: launch is scheduled
for April,~2018 and the orbit insertion for December,~2024. The nominal
duration of the whole mission is 1 year, with a possible extension to
2 years.

The Mercury Orbiter Radioscience Experiment (MORE) is one of the on
board experiments whose goals are:
\begin{itemize}
\item[a)] to determine the gravity field of Mercury and its
rotation state (gravimetry and rotation experiments); 
\item[b)] to study possible violations of the general relativity (GR)
theory of gravitation (relativity experiment);
\item[c)] to provide the spacecraft position for geodesy experiments;
\item[d)] to contribute to planetary ephemerides improvement.
\end{itemize}

Such precise experiments are possible thanks to a multi-frequency
radio link (in $X$ and $K_a$~bands) allowing to eliminate the uncertainty
in the refraction index due to plasma content along the radio waves
path \cite{IB}. The MORE experiment provides the necessary $K_a$-band
transponder and the system to compare the delays in a 5-way link, in
combination with instruments installed at the ground stations.

Orders of magnitude for the accuracy which can be achieved in this way
are $3$ micron/s in range-rate (two-way, at 1000~s of integration time) and 30~cm in range (two-way,
at 300~s of integration time): the relative accuracy in range is better than $10^{-12}$. This implies the signal
to noise ratio (S/N) of all the relativistic effects (both in the
dynamics and in the observation equations) is very large, in
particular for the range measurements. By using the nonlinear least squares method, range and range-rate data 
will be fitted to an accurate theoretical model to estimate the physical 
parameters relative to MORE as well as their uncertainties.

One of the most ambitious purposes of the mission is to 
attempt to constrain the Nordtvedt parameter $\eta$ describing violations 
of the strong equivalence principle (SEP). The equivalence principle (EP) states the equality between inertial and  gravitational mass. As a consequence, there is the universality 
of the free fall that allows the geometrical description of the gravity in  GR. 
So far, there is no experimental disproof of it. In its weak form the EP states that strong and electro-weak interactions do not influence the falling of a body, with a negligible self-gravity, 
in an external gravity field. This is called the weak equivalence principle (WEP).
The strong form extends the validity of the EP to bodies with measurable self gravitational energy, like stars or planets.
The inequality between inertial mass $m_i^I$ and gravitational mass $m_i^G$ of a body can be parametrized as \cite{milani2002,damour1996,congedo2016}
\be
 m_i^G = m_i^I (1+ \delta_i + \eta\, \Omega_i),
\label{mg_mi}
\ee
where $\Omega_i$ is the ratio between self-gravitational and rest energy of the body.

The validity of the WEP corresponds to $\delta_i = 0$ and $\Omega_i=0$. On the other hand, when $\Omega_i\neq 0$, if both $\delta_i$ and
$\eta$ are zero the SEP is valid.
The values of $\Omega_i$ for the Sun, Earth and Moon are, respectively $\Omega_s=-3.52 \times 10^{-6}$; 
 $\Omega_e=-4.64 \times 10^{-10}$; $\Omega_m=-1.88 \times 10^{-11}$ \cite{williams2009}.\\
 Since Newton, EP has been tested several times by laboratory experiments, mainly by  torsion balances, but 
 laboratory objects have a very small self-gravity ($\Omega_i \lessapprox 10^{-26}$), therefore no information about SEP can arise from ground based experiments only. 
 Hence,  an EP test is in practice  a WEP test.\\
The first test of the EP involving celestial bodies (Earth, Moon and Sun) has been proposed by Nordtvedt \cite{nordtvedt1968}: in case of EP
violation the different rate of free falling of Earth and Moon towards the Sun generates a signal in the Earth-Moon range. 
This signal carries information about both WEP and SEP since Earth and Moon have different gravitational self-energies but also compositions.
Regarding the SEP violation, the amplitude of the signal on Earth-Moon distance is proportional to $(\Omega_e-\Omega_m)$, at the Moon's synodic frequency, and it has  
been estimated to be $\approx 13\, \eta$ meters (the so-called Nordtvedt effect).
In order to separate weak and strong EP, laboratory results  involving test bodies with chemical composition similar to that of
Earth and Moon have been used. 
The precision achieved on WEP measurements is currently $\sigma[\delta a_{WEP}/a]=1.4 \times 10^{-13}$ \cite{adelberger2009}, but this result is expected to be 
 improved by 2 orders of magnitude thanks to the recently launched ESA mission MICROSCOPE  \cite{touboul2012}.
The Earth-Moon distance has been measured with increasing precision in the last 46 years by lunar laser ranging (LLR) \cite{murphy2012} and 
 the precision on Earth-Moon relative differential accelerations is currently $\sigma[\delta a_{em}/a_\odot ]=1.3 \times 10^{-13}$ 
\cite{williams2009}. This result is comparable to that achieved by ground experiments.\\
In order to estimate $\eta$ is necessary to consider both experiments and the resulting RMS is $\sigma [\eta]= 4.4 \times10^{-4}$ \cite{adelberger2009, williams2009, turyshev2007}.\\
SEP violation can be tested also by planetary ranging, i.e. by radio-tracking between Earth and an object orbiting around the Sun (a spacecraft or another planet).
The advantages with respect to the Earth-Moon tests are: 
$i$) the signal is proportional to $\Omega_s$ which is 4~orders of magnitude larger than $\Omega_e- \Omega_m$ and $ii$) the baseline is in general larger \cite{turyshev2004}.\\
A test for the SEP violation by ranging measurements between Earth and $L_1$ or $L_2$ Earth-Sun Lagrangian points has been recently proposed by \cite{congedo2016}, while 
 the same experiment by Earth-Mars or Earth-Phobos ranging has been described by \cite{turyshev2004,turyshev2007,turyshev2010}.\\
Thanks to precise Earth-Mercury range measurements, the BC mission will provide data to estimate with good accuracy $\eta$ and other
  post-Newtonian (PN) parameters \cite{milani2002, MTVLC}.
Regarding the EP, the expected precision for relative Earth-Mercury acceleration is $\sigma [\delta a_{earth-merc.} /a_\odot]\approx 10^{-11}$.
It is 2 orders  of magnitude worse than the precision achieved by laboratory tests, therefore the EP violation test of BC  is in practice a SEP test. 
Moreover, the SEP violation signal is proportional to $\Omega_s$, therefore the parameter $\eta$ can be estimated with high precision.

The purpose of this work is to quantify $\sigma[\eta]$ in presence of systematic effects due to planets and asteroids.
To do this, we will use both a numerical approach (by using the software {\em ORBIT14}, described in Section~\ref{sec:models}) 
 and an analytical one. This latter being necessary to take into account the effects of the 
 experimental uncertainties of gravitational parameters ($\mu_i=G m_i^G$, hereafter GPs) of planets and asteroids.
 
This paper has the following structure. 
In Section~\ref{sec:models} we are going to give a brief description of the
mathematical models and of the structure of the simulation software
used and we will present the result of a nominal simulation of
the relativity experiment. 
Section~\ref{sec:sep} is devoted to the detailed description of the experiment to estimate $\eta$ and 
of the analytical method followed. 
Finally, in Section~\ref{sec:concl} we will draw some conclusions.


\section{Mathematical models and software structure}\label{sec:models}

In this section we briefly describe the mathematical models that are the basis of the software {\em ORBIT14} built to process the data of MORE  and used as simulation
 software to infer some results about the estimation of parameters. {\em ORBIT14}, has been developed in the last six years by the Celestial Mechanics 
 Group of the University of Pisa and it has the capability to simulate the {\em relativity experiment}, the {\em gravimetry and rotation
experiment} \cite{MOREgrav} of the BC mission, and also the Radioscience experiment of the NASA mission JUNO \cite{TDSM}.

Concerning the relativity experiment, we need to solve an orbit determination problem with a full relativistic model (also for
the observable computations, see \cite{TMV}), including the terms expressing the violations of general relativity with the
PN parameters, such as $\gamma, \beta, \eta, \alpha_1, \alpha_2$. 

The equations of motion of Mercury and of the Earth-Moon barycenter
(EMB) have been implemented using the parametric post-Newtonian
approach: they are linearized with respect to the small parameters
$v_i^2/c^2$ and $G\,m_i/(c^2\,r_{ik})$, where $v_i$ is the barycentric
velocity for each of the bodies of mass $m_i$, $c$ the speed of light
and $r_{ik}$ a mutual distance, appearing in the metric of the curved
space-time, hence in the equations for geodesic motion.  This can be
done adding to the Lagrangian of the N-body problem some
corrective terms of PN order 1 in the small parameters (as described
in \cite{MTVLC}).\\
Here we are interested in the model for the violation of SEP. We can
consider that there are for each body $i$ two quantities $\mu_i=G m_i^G$ and
$\mu^I_i=G m_i^I$, one in the gravitational potential (including the
relativistic part) and the other in the kinetic energy, not considering the relativistic
correction for the masses. If there is a violation of the SEP involving body $i$, with
a fraction $\Omega_i$ of its mass due to gravitational self-energy,
then from \eref{mg_mi} and neglecting WEP
\be
\mu_i= \mu^I_i\,(1+\eta \, \Omega_i)\Longleftrightarrow \mu^I_i= \mu_i\,(1-\eta \, \Omega_i) + {\cal O}(\eta^2) \,\, .
\ee
Neglecting ${\cal O}(\eta^2)$
terms this is expressed by a Lagrangian term $\eta\, L_\eta$, with an
effect on body $i$:
\be
L_\eta= -\frac 12\;\sum_{i}\, \Omega_i\, \mu_i\, v_i^2\ \  \Longrightarrow \ \ 
\dersec{\rrr{i}}{t}=(1+ \eta\, \Omega_i) \left.\dersec{\rrr{i}}{t}\right|_{\eta=0}\;
\ .
\ee
The largest effect of $\eta$ is a change in the center of mass integral (where dots stand for terms of order $2$~PN and higher, including the neglected terms in the masses)
\be
\PP=\sum_j \derparz{L}{\vvv{j}}=
\sum_j\,(1-\eta\,\Omega_i)\,\mu_i\,\vvv{j}+ \ldots \eqspa
\dertot{\PP}t=\zero
\ee
and if the center of mass is the origin, the position of the Sun has to be corrected  (to indicate the Sun we use the subscript "0")
\be
\ve r_0=\frac{-1}{\mu_0\, (1-\eta\,\Omega_0)}\sum_{j\neq
0}(1-\eta\Omega_j)\, \mu_j\,\rrr{j}+\ldots\ .
\label{eq:bar}
\ee
The partial derivative of the acceleration of the body $j$ with respect to $\eta$ is
\be
\derparz{}{\eta}\dersec{\rrr{j}}{t}=
\Omega_j \left[\frac{\mu_0}{r^3_{j0}}\,\rrr{j0} + \sum_{i\neq j,0}
\frac{\mu_i}{r^3_{ji}}\rrr{ji}\right]+\derparz{\ve r_0}{\eta} 
\derparz{(\mu_0/r^3_{j0})}{\ve r_0}
\, ,
\ee
where the first term is the direct, the second the indirect $\eta$-perturbation, and 
\be
\derparz{\ve r_0}{\eta}=\sum_{i\neq 0}\,(\Omega_j-\Omega_0)\,
\frac{\mu_i}{\mu_0}\,\rrr{i} \,\, .
\ee
By combining together and omitting smaller terms with
$\Omega_i\,\mu_k$ (with $i, k\neq 0$) or ${\cal O}(\eta^2)$ 
\be
\label{eq:ddeta}
\derparz{}{\eta}\dersec{\rrr{j}}{t}=\Omega_j\,\mu_0\;
\frac{\rrr{j0}}{r^3_{j0}}
-\Omega_0\, \derparz{(1/r^3_{j0})}{\rrr{0}}\,
\sum_{i\neq 0}\,\mu_i\,\rrr{i} \,\, ,
\ee
with a direct (small parameter $\Omega_j\,\mu_0$) and an indirect
(small parameter $\Omega_0\,\mu_i$) part.


\subsection{ORBIT14 software structure}\label{sec:o14}
Since the real data from the spacecraft will be available only in
2025, the first program of the software is the {\it
  simulator}, which generates fictitious sets of observables,
non-gravitational accelerations and initial orbital elements for the
probe at the central time of each observed arc. These are obtained by
propagating the orbit of the spacecraft starting from some initial
conditions taken by the spice kernel generated by the navigation team
of the mission.

The core of the orbit determination software is the {\it corrector},
the purpose of which is to estimate the parameters we are interested
in. This program is the one that will be used to analyze real
data. The corrector follows a classical approach (see, for instance,
\citep{MG}), and its aim is to perform a non linear least squares fit
to compute a set of parameters $\mathbf{q}^*$ which minimizes the
target function
\begin{equation}\label{eq:1}
 Q(\mathbf{q})=\frac{1}{m}\boldsymbol{\xi}^T(\mathbf{q})
\mathbf W\boldsymbol{\xi}(\mathbf{q})=\frac{1}{m}\sum_{i=1}^mw_i\xi_i^2(\mathbf{q}),
\end{equation}
where $m$ is the number of observations and
$\boldsymbol{\xi}=\mathcal{O}-\mathcal{C}$ is the vector of {\it
  residuals}, difference between the observed quantities $\mathcal{O}$
and the predicted ones $\mathcal{C}(\mathbf{q})$, computed using
suitable models and assumptions. In our case, the observed quantities
are range and range-rate data, while the computed observables are the
results of the light-time computation (see \cite{TMV} for more
details) as a function of all the quantities $\mathbf{q}$ we want to
estimate ($w_i$ is the weight associated to the $i-$observation).

The procedure to compute $\mathbf{q}^{*}$ is based on a modified
Newton's method known in the literature as \emph{differential
corrections method}; see e.g. \cite{MG}. Let us define
\be
\mathbf B={\partial\boldsymbol{\xi}\over\partial\mathbf{q}}(\mathbf{q}),
\quad \quad \mathbf C=\mathbf B^\mathbf T \mathbf W \mathbf B,
\ee 
which are called the {\it design} matrix and
the {\it normal} matrix, respectively. Then the correction:
\begin{equation}\label{eq:2}
\Delta \mathbf{q}=\mathbf C^{-1} \mathbf D \quad \rm{with   }\, \mathbf D=-\mathbf B^T \mathbf W\boldsymbol{\xi}
\end{equation}
is applied iteratively until either $ Q$ does not change meaningfully
from one iteration to the other or $\Delta \mathbf{q}$ becomes smaller
than a given tolerance.

Concerning the observations, we suppose to have two ground stations,
one observing in $K_a$-band at the Goldstone Deep Space Communications Complex in California
(U.S.) and the other in $X$-band at the Cebreros station in Spain. With this
scenario, the observations are split in arcs, with interruptions of
tracking not exceeding one hour, namely the {\it observed arcs}, with
a duration from 14 to 19 hours. The arcs are separated by intervals in
the dark lasting from 5 to 10 hours. We call {\it extended arc} an
observed arc broadened out from half of the dark period before it to
half of the dark period after it.

In order to estimate the parameters we do not use a classical
multi-arc strategy\footnote{Every single arc has its own set of
  initial conditions; in this way, the actual errors in the orbit
  propagation, due to lack of knowledge in the non-gravitational
  dynamical model, can be reduced by an over-parametrization of the
  initial conditions.}, as described for example in \cite{MG}, but a
{\bf constrained multi-arc strategy}. This method is established on
the idea that each observed arc belongs to the same object (the
spacecraft) and thus the orbits corresponding to two subsequent
extended arcs should coincide at the connection time in the middle of the
non-observed interval (see \cite{ACMT} for more details).

\subsection{Nominal simulation}\label{sec:nominal}
The simulation scenario (the same as described in \cite{SchettinoIEEE}) consists of a 365 arcs long simulation, which corresponds to about one year, starting on March, 27th~2025. 
The main assumptions made are briefly described as follows:
\begin{itemize}
\item two ground stations are available for tracking, one at Goldstone Deep Space Communications Complex (California, USA) for the $K_a$-band and the other in Spain
, at Cebreros station, for $X$-band; range measurements are taken every 120~s and range-rate every 30~s, both with top accuracies;
\item we impose the Nordvedt equation \cite{nordtvedt1970}, i.e. we assume a metric theory to remove the approximate symmetry between $\beta$ and $J_{2\odot}$: 
\be
\eta=4(\beta -1)-(\gamma -1)-\alpha_1 -\frac{2}{3}\alpha_2
\ee
\end{itemize}
The MORE relativity experiment consists in solving for the following parameters: the PN and related parameters 
($\gamma$, $\beta$, $\eta$, $\alpha_1$, $\alpha_2$, $J_{2\odot}$, $\mu$, $\zeta$) plus $6$ initial conditions for 
Mercury barycenter $\{X_m,Y_m,Z_m,\dot{X}_m,\dot{Y}_m,\dot{Z}_m\}$ and 6 for the EMB 
$\{X_e,Y_e,Z_e,\dot{X}_e,\dot{Y}_e,\dot{Z}_e\}$ with respect to the SSB in the 
Ecliptic J2000 reference frame\footnote{Since, from a relativistic point of view, we are not dealing with a generic 
space-time but with the one where we are ``now'', we need to solve for the initial conditions of the bodies which 
affect the measurements, i.e. Mercury and EMB. The other planetary ephemerides are taken by JPL ones}. 
Due to rank deficiency, among the $12$ parameters (positions and velocities), only $8$ can be determined simultaneously. 
They are position and velocity of Mercury and two components of the velocity of the EMB.\\ 
We solve for all the parameters listed above in a global least square fit. We performed both an analysis based
 on formal statistics (standard deviations and correlations) as given from the formal covariance matrix 
 $\mathbf{\Gamma}=\mathbf{C}^{-1}$, and an analysis based on ``true'' errors, defined as the difference 
 between the value of parameters at convergence and the simulated value. We perform a statistical 
 analysis over 10 runs, each time varying the random generator of Gaussian distribution and we consider 
 as true error the distribution mean value. Concerning the convergence requirements, we imposed a tolerance 
 treshold in target function variation between two subsequent iteration of $10^{-4}$ and we verified that the differential correction process reached this condition always in 6 iterations.\\
The results for PN and related parameters in terms of formal uncertainty are shown in Table \ref{tab_relnom}.

\begin{table}[!t]
\begin{tabular}{ll}
\hline
\hline

 parameter      \hspace{1cm}           &               \bf  Formal sigma \\[4pt]
                                   \hline 
$\beta$                                       &              $6.21\times 10^{-7}$ \\   
$\gamma$                                  &             $7.65\times 10^{-7}$  \\ 
$\eta$                                         &              $1.93\times 10^{-6}$  \\ 
$\alpha_1$                                  &             $4.5\times 10^{-7}$  \\ 
$\alpha_2$                                  &             $7.6\times 10^{-8}$  \\  
$\mu_0$                                     &             $3.9\times 10^{13}$  \\ 
$J_{2\odot}$                               &             $3.8\times 10^{-10}$ \\  
$\zeta=\dot \mu_0/\mu_0$          &             $2.0\times 10^{-14}$ \\  
\hline
\end{tabular}
\caption{Results for PN and related parameters estimation in the "ideal" case (planetary masses have no errors). Units are [cm$^3$/s$^2$] for $\mu_0$ and [yr$^{-1}$] for $\zeta$ .}
\label{tab_relnom}
\end{table}
                                         
The parameters $\eta$ and $\zeta$ present true errors always higher than formal ones because they are very sensitive to the effect of systematic errors in range. 
Nevertheless, the expected results of MORE could hence improve the actual knowledge, 
even if there are some intrinsic problems due to the uncertainties in the masses and ephemerides of solar system bodies (see the next section and \cite{tommei2015}).


 \section{Analytical model and sources of uncertainties}\label{sec:sep}
 
  {\em ORBIT14} integrates the orbits of EMB and Mercury, while the trajectories of planets and asteroids are taken from JPL
  ephemerides. The position of the Sun is obtained from \eref{eq:bar} as a function of positions and (relativistic) masses of the other bodies.
  
  We want to test  how the spurious signal due to a wrong value of the mass of a planet affects the estimation of $\eta$.
This kind of test cannot be performed by {\em ORBIT14}. In fact, we could try to fit the simulated data by using a model with a slightly different value of the mass.
At the first iteration, the modeled barycentric orbits of the planets are the same of the data, since they come from ephemerides, 
but the position of the Sun is slightly different due to the different mass of the planet.
  Hence,  all mutual distances among solar system bodies are altered and the modeled MPO-Earth range turns to be 
  very different from the simulated data. This implies an unphysical systematic effect on the parameters estimation.
  The parameter $\eta$ heavily feels this effect because it enters into the relativistic equation of the center of mass \eref{eq:bar}.
  
 To perform this test,  we develop a  heliocentric analytical model and we include in our calculus the parameters whose 
 signals are expected to be correlated with the SEP violation signature.\\
 In order to avoid systematic effects, GPs must be added to the set of parameters to be fitted and their errors must
  be taken into account in terms of constraints (hereafter {\em apriori}) to be included to the global covariance analysis. \\
All signals involved in the relativity experiment have  frequencies  of the same order of planetary mean motions. 
For this reason, we can neglect the motion of MPO around Mercury (the orbital period is approximately 2~hrs)  and we will consider the Mercury-Earth range.\\
Current uncertainties of planetary GPs go from $2.8\times 10^{-4}$ (Mars) to $\approx 10.5$~[km$^3$/s$^2$]  (Neptune) \cite{luzum2011}. 
Regarding asteroids, relative errors can be very large (50\% or more).\\
 To summarize, we will calculate the signatures on the  Earth-Mercury range due to:
 \begin{enumerate}
  \item initial conditions of Earth and Mercury,
  \item SEP violation  (parameter: $\eta$),
 \item planets/asteroids (parameters to be fitted: GPs),
 \item secular variation of Sun's GP $\mu_0$. Parameters to be fitted are $\delta_{\mu_0}$ (displacement from the nominal GP of the Sun at the starting epoch) and its rate of change in time $\zeta=\dot \mu_0/\mu_0$,
 \item PN parameter $\bar \beta=\beta-1$,
 \item Sun's quadrupole coefficient $J_{2 \odot}$.
\end{enumerate}
 The parameter $\gamma$, which is related to the curvature produced by unit rest mass, for simplicity has not been considered. 
 However, this is not reductive since the best estimation of $\gamma$ is expected to be given after the dedicated superior conjunction experiment 
  (SCE) and its RMS $\sigma[\gamma]=2.0\times 10^{-6}$ will be inserted as an {\em apriori} into the Nordtvedt equation.\\
 
\subsection{Analytical model}
We adopt the notation of \cite{moyer2000}: we define $\ve r_{ij}=\ve r_j-\ve r_i$ and $r_{ij}=\vert \vert \ve r_{ij}\vert \vert$ where  $\ve r_i$ is the coordinate of planet $i$ in an inertial reference frame.\\
Planets are numbered from 1 (Mercury) to 8 (Neptune), while 0 is referred to the Sun.\\
We will use subscripts/superscripts $i$ and $k$ to indicate Mercury and Earth, respectively, while $j$ will be used for an arbitrary {\em perturber} body (planet or asteroid).\\
We call $\ve q =\{q_1,..., q_N\}$ the set of parameters to be fitted and $\rho_{13}(t,\ve q)$ the analytical model of the Mercury-Earth range to be calculated as a function of $\ve q$.\\
We will describe the motion of the planet $i$ as a small perturbation from a heliocentric circular orbit with radius $R_{0i}$ equal to the semimajor axis and  mean motion $n_i=\left[(\mu_0+\mu_i)/R_{0i}^3 \right]^{1/2}$.\\
 { For all parameters except $J_{2\odot}$ (see below)  inclinations will be neglected, while eccentricities are assumed to be zero in all cases}. \\
The displacement from the reference orbit is $\delta \ve r_i= x_i \ve u_r^i+ y_i \ve u_t^i+z_i \ve u_z^i$ where $\ve u_r^i,\ve u_t^i, \ve u_z^i$ are  radial, along-track and out-of-plane unit vectors, respectively.\\
We will express the position of $k$ relative to $i$ as
\be
\ve r_{ik}= \ve R_{ik}+\sum_{m=1}^N q_m \delta \ve  r_{ik,m}
\ee
where $\ve R_{ik}=R_{0k}\ve u_r^k- R_{0i} \ve u_r^i$.\\
Since terms in the summation are small, at the first order the range is
\be
 \rho_{ik}= \vert \vert \ve r_{ik} \vert \vert \approx R_{ik}+ \sum_n q_n \frac{\delta \ve r_{ik,n} \cdot \ve R_{ik}}{R_{ik}}\, .
\label{eq:pert1}
 \ee
The factor $1/R_{ik}$ can be expressed by the Legendre polynomials $P_n$ (for $R_{0i}<R_{0k}$) \cite{ashby2007}
\be
 \frac{1}{R_{ik}}= \frac{1}{R_{0k}}\sum_{n=0}^\infty \left( \frac{R_{0i}}{R_{0k}}\right)^n P_n (\cos \Phi_{ik})
\label{eq:legendre}
 \ee
where we defined $\Phi_i=n_i t+\varphi_i$ and $\Phi_{ik}=(n_k-n_i)\,t+\varphi_k-\varphi_i$.\\
Afterwards,  for each $q_n$ we will calculate the corresponding $\delta \ve r_{ik,m}$.\\
We decompose the perturbation on $i$ (or $k$) as a sum of radial and along-track forces $\sum_n q_n' (R_n^i \ve u_r^i+T_n^i \ve u_t^i)$, where $\ve q'$ represents the subset  
of $N-12$ {\em dynamical} parameters to be estimated (all but  initial conditions of Earth and Mercury). For simplicity we assume that perturbations are on the ecliptic plane.\\
Since we are assuming that $\vert \vert \delta \ve r_i \vert \vert \ll R_{0i}$, we can use the first order Hill's equations \cite{cw1960}. They are
\be
\label{eq:hill}
\begin{split}
\ddot x_i - 2\,n_i \dot y_i -3 \,n_i^2 x_i &=\sum_n q_n' R_n^i\\
\ddot y_i+2 \,n_i \dot x_i  &= \sum_n q_n' T_n^i\\
\ddot z_i+ n_i^2 \dot z_i  &= 0.
 \end{split}
 \ee
Solutions are the sum of the homogeneous part $\{\hat x_i,\hat y_i,\hat z_i\}$ plus the contributions $\{x_i',y_i',z_i'\}$ due to the perturbing forces. Since \eref{eq:hill} are linear, the inhomogeneous terms  can be calculated once at time and finally summed together.

\subsubsection{Initial conditions}
Here we calculate the signature on range due to initial conditions of Earth and Mercury. 
We  consider two cases: initial conditions expressed in heliocentric and in barycentric (SSB) reference frame. 
In this latter case and additional signal which depends on $\eta$ must be included.\\
Referring to \eref{eq:hill}, we express the state vector $\ve v=\{x_i,y_i,z_i,\dot x_i, \dot y_i, \dot z_i \}$ of body $i$ as
\be
\ve v=\ve {\hat v} \ve A +   \ve v' \ve q'
\ee
where the first term on the right side represents the homogeneous solution of \eref{eq:hill}. It is the product of a 6x6 matrix $\ve {\hat v}$ [see \eref{eq:hatv}  in 
Appendix] and the vector $\ve A$ of 6 coefficients to be fixed by the initial conditions.\\
The second term, the inhomogeneous solution, is the product of the $6 \times (N-12)$ matrix $\ve v'$ of the particular solutions and the column vector $\ve q'$ of the parameters~$q'_n$.\\
We pass to the coordinate system with fixed axes, and we express the state vector as  $\ve v= {\cal R} \,\ve \delta \ve x$, where $\cal R$ is the corresponding  rotation matrix [see \eref{eq:R}]. \\
We rewrite $\ve A$ in terms of $\delta \ve x $ and $\cal R$ at $t=0$ (say $\delta \ve x_0,  {\cal R}_0$) and,  
defining ${\cal V}= \ve {\hat v}\,  \ve {\hat v}_0^{-1}$ [see \eref{eq:calV}], we get
\be
\label{eq:v}
\ve v=  {\cal V} \, {\cal R}_0\, \delta \ve x_0 +  \left(\ve v' -{\cal V}\, \ve v_{0}' \right) \ve q'\,.
\ee
The first term represents the "signal" due to initial conditions $\delta  \ve x_0$, while the signal due to parameter $q_n'$ (the term into brackets) is the sum of the particular solution $\ve v'$ 
plus the homogeneous solution corresponding to $\delta \ve x_0=0$.\\
The complete set of parameters to be determined by the ranging between $i$ and $k$ is
\be
\ve q=\{\delta \ve x_0^i,\delta \ve x_0^k, \ve q'\}\,.
\ee
Defining 6x6 matrix $f_{\alpha \mu}^i$ and $6\times(N-12)$ matrix $g_{\alpha \mu}^i$ as
\be
f_{\alpha \mu}^i={\cal R}^{-1} \ve  {\cal V}\ve {\cal R}_0 ; \qquad g_{\alpha \mu}^i={\cal R}^{-1} (\ve v' -{\cal V}\, \ve v_{0}' )
\ee
($\cal R, \cal V$ and $\ve v$ are referred to body $i$) and  
using $\alpha=1, 2 , 3$ to indicate the spatial components in the fixed axes coordinate system  of $\delta \ve r_{ik,m}$, we get 
 \be
\delta \ve r_{ik,m}= 
 \begin{cases}
- f_{\alpha \lambda }^i   &                            \quad \lambda=m      \mbox{ and } m \leq 6 ; \\
  f_{\alpha \lambda}^k   &                            \quad \lambda=m-6   \mbox{ and } 7\leq  m \leq 12 ;\\
 g_{\alpha \lambda}^k-g_{\alpha \lambda}^i &        \quad \lambda=m-12  \mbox{ and }    m >12.
 \end{cases}
 \label{eq:pert2}
 \ee
Finally, by Eqs.~(\ref{eq:pert1}, \ref{eq:pert2}) we  obtain the perturbation on range due to each element of $\ve q$.\\
The barycentric initial state vector $\ve X_0^i$ is related to the heliocentric one by
\be
\label{eq:X0}
 \ve X_0^i = (1+\eta \Omega_0 ) \ve R_0 +R_{0i} \ve s_i+\ve \delta \ve x_0^i 
\ee
where
\be
\ve s_j=\{\cos \varphi_j,\sin \varphi_j,0,-n_j \sin \varphi_i, n_j \cos \varphi_i,0\}
\ee
and
\be
\ve R_0 = -\frac{\sum_{j \neq 0} \mu_j  R_{0j}\ve s_j}{\sum_j \mu_j} 
\ee
is the position of the Sun with respect to the SSB in the case $\eta=0$.
Therefore, if we pass to  barycentric initial conditions we must take into account additional signals due to $\mu_j$s and  $\eta$.\\
By \eref{eq:X0} we can express \eref{eq:v} as a function of $\ve X_0^i$. Adopting now $\ve q=\{\ve X_0^i, \ve X_0^k, \ve q' \}$, 
we calculate the extra signals due to $\eta$ and $\mu_l$ to be added to $\delta \ve r_{ik,m}$ in \eref{eq:pert2}.
They  are reported in \eref{eq:additional}.

\subsubsection{SEP violation and planets/asteroids  contributions}\label{subsect_sep}
In the heliocentric reference frame the equations of motion of a planet $i$, in the case $\eta \neq 0$, are
\cite{anderson1996,milani2002,turyshev2004,ashby2007,congedo2016}
\be
\label{eqr0i}
\ddot { \ve r}_{0i} = -\dfrac{\mu^\star}{r_{0i}^3}\ve r_{0i} + \displaystyle \sum_{j \neq i \neq 0} \mu_j \left[(1+ \eta \, \Omega_i)\dfrac{\ve r_{ij}}{r_{ij}^3} - (1+\eta\, \Omega_0) \dfrac{\ve r_{0j}}{r_{0j}^3}\right]  ,
\ee
where the summation is extended to all solar system bodies and 
\be
\mu^\star=\mu_0+\mu_i+\eta (\mu_i \Omega_0+\mu_0 \Omega_i)\,.
\ee
From \eref{eqr0i} a high correlation among planetary perturbations (depending on $\mu_j$) and SEP violation is evident.\\
We separate the contributions of parameters $\eta$ and $\mu_j$ and we project them on radial and along-track directions. 
Since $\Omega_i \ll \Omega_0$, the SEP violation contribution can be simplified \cite{ashby2007,tommei2015,congedo2016}
\be
\label{eta_approx}
\begin{split}
R_\eta^i &\approx  - \Omega_i n_i^2 R_{0i} -\Omega_0 \displaystyle \sum_{j\neq i \neq 0}  \dfrac{\cos \Phi_{ij}}{R_{0j}^2}\, ;\\
T_\eta^i & \approx -\Omega_0 \displaystyle \sum_{j\neq i \neq 0}  \dfrac{\sin \Phi_{ij}}{R_{0j}^2} \,;
\end{split}
\ee
[see  \eref{eta_exact} for the complete expression].\\
The signal  contains a small permanent radial displacement due to a ''direct'' term $\propto \Omega_i$ and an ''indirect'' term, which depend on $\Omega_0$. 
These terms have been calculated in the SSB frame in \sref{sec:models} [see \eref{eq:ddeta}].\\
The particular solution of \eref{eq:hill} relative to parameter $\eta$  can be written as $\{x_{i,\eta}',y_{i,\eta}'\}$ where
\be
  \label{eq:part_eta}
\begin{split}
 x_{i,\eta}' & = \Omega_i\,\dfrac{R_{0i}}{3} +\Omega_0 \sum_{j \neq i}  \dfrac{\mu_j}{R_{0j}^2} \dfrac{1 + 2 \,n_i/n_{ji}}{\,n_{ji}^2-n_i^2} \cos \Phi_{ji},\\
 y_{i,\eta}' &=-\Omega_0 \sum_{j \neq i}  \dfrac{\mu_j}{R_{0j}^2}\,\dfrac{1+2 \,n_i/n_{ji} + 3 \,n_i^2/n_{ji}^2}{\,n_{ji}^2-n_i^2} \sin \Phi_{ji}.
 \end{split}
 \ee
Similarly, perturbations on planet $i$ due to body $j$ are (radial and along-track) \cite{ashby2007}
\be
\label{RTplanetj}
\begin{split}
R_{\mu_j}^i&=  \displaystyle \sum_{j\neq i \neq 0}  \left(   \dfrac {R_{0j}\cos \Phi_{ij}-R_{0i}}{R_{ij}^3}-  \dfrac{\cos \Phi_{ij}}{R_{0j}^2}\right)\, ,\\
T_{\mu_j}^i &=    \displaystyle \sum_{j\neq i \neq 0}   \left( \dfrac {R_{0j}}{R_{ij}^3}-  \dfrac{1}{R_{0j}^2}\right)  \sin \Phi_{ij}\, .
\end{split}
\ee
The coefficient $1/R_{ij}^3$ can be calculated  from \eref{eq:legendre} and  expressed as a Fourier cosine series with fundamental frequency $\Phi_{ij}$ [see \eref{eq:r3}]. 
Therefore, we can write 
\be
\begin{split}
R_{\mu_j}^i &=   \displaystyle\sum_{l=0}^\infty \,a_{j,l}  \cos ( l\, \Phi_{ij});\\
  T_{\mu_j}^i &=   \displaystyle\sum_{l=1}^\infty \,b_{j,l}  \sin ( l\, \Phi_{ij});
\label{eq:plan}
\end{split}
\ee
 and  coefficients $a_{j,l}$ and $b_{j,l}$ are reported in \eref{eq:ajbj}.\\
The radial and along-track components of  $\{x_{i,\mu_j}',y_{i,\mu_j}'\}$  are, respectively 
\be
\begin{split}
  x_{i,\mu_j}' &= - \dfrac{a_{j0}}{3\, n_i^2}- \displaystyle\sum_{l=1}^\infty \dfrac{a_{j,l} - 2 b_{j,l}/(l n_{ji})}{l^2 n_{ji}^2-n_i^2} \cos \Phi_{ji}\, ;\\[12pt]
  y_{i,\mu_j}' &=- \displaystyle\sum_{l=1}^\infty \dfrac{b_{j,l}-2 a_{j,l} n_i/(l n_{ji})+3 b_{j,l}n_i^2/(l^2 n_{ji}^2)}{l^2 n_{ji}^2-n_i^2}  \sin \Phi_{ji} \, .
 \end{split}
 \label{eq:part_muj}
 \ee
Finally, by applying \eref{eq:pert1} and \eref{eq:pert2}, the Earth-Mercury range perturbations due to parameters $\eta$ can be written as
\be
\label{eq:eta}
\begin{split}
 \delta \rho_{13}^\eta &=  \sum_{l=1}^\infty {\cal D}_l \cos \,( l\, \Phi_{13} )+ \\
 &+ \sum_{\substack{j=pl.\\+ast.}} \,\, \sum_{\substack{p,q,r\\  \in \mathbb{Z}}} {\cal I}_j^{pqr} \,\cos \,(p\,\Phi_1+q\, \Phi_3+r\, \Phi_j) 
 \end{split}
\ee
where coefficients ${\cal D}_l$ are due to the "direct effect" and they depend only on  $\Omega_1$ and $\Omega_3$.\\
For all perturbing bodies considered, we calculated the numerical values of ${\cal D}$ and ${\cal I}$ using the complete expression \eref{eta_exact}.
They are reported (for planets from Mars to Neptune) in \tref{tab:eta}.\\
An analog expression  can be written for the perturbation due to body $j$
\be
 \delta \rho_{13}^{\mu_j}=  \sum_{\substack{p,q,r\\
 \in \mathbb{Z}}} {\cal J}^{pqr} \,\cos \,(p\,\Phi_1+q\, \Phi_3+r\, \Phi_j).
\label{eq:muj}
\ee
Numerical coefficients for ${\cal J}$ are reported in \tref{tab:pla}.

\subsubsection{Range signature due to a variation of Sun's GP}
It is well known that the GP of the Sun is not constant in time due to Sun's mass loss and to a possible (but  unconfirmed) dependence of $G$ on time.\\
Matematically this corresponds to
\be
 G M_\odot (t) = \mu_0    \left[1 + \zeta (t-t_0)\right]+\delta_{\mu_0}
\ee
where $t_0$ is the epoch (hereafter $t_0=0$) when the GP of the Sun is equal to $\mu_0+\delta_{\mu_0}$. The small parameter $\delta_{\mu_0}$ has been 
introduced to account that  true and  nominal value $\mu_0$ of the GP of the Sun are not the same. The parameter $\zeta=\dot \mu_0/\mu_0$ is  a small (constant) rate of change.\\
If $\zeta \neq 0$ and/or $\mu_0$ is slightly different from the nominal value, a radial perturbation will be present. The Hill's equations are 
\be
\begin{split}
 \ddot x_i-2 n_i \dot y_i - 3 n_i^2 x_i& = -\dfrac{ \delta_{\mu_0}+ \mu_0 \zeta \,t }{R_{0i}^3}\,,\\[8pt]
 \ddot y_i+2 n_i \dot x_i&=0\,;
\end{split}
\ee
and a particular solution is
\be
x_i'= -\dfrac{\delta_{\mu_0}+ \mu_0 \zeta \,t }{n_i^2 R_{0i}^2}; \qquad y_i'= \frac{\mu_0 \zeta t^2+ 2 \delta_{\mu_0} t}{n_i R_{0i}^2}\,.
\label{eq:part_zeta}
\ee
By  Eqs.~(\ref{eq:pert1}), (\ref{eq:pert2})  we obtain the range signature due to $\zeta$ and $\delta_{\mu_0}$.

\subsubsection{Range signature due to $\beta \neq 1$}
The PN parameter $\beta$ is related to the nonlinearity in the superposition of gravity. In GR is, by definition, $\beta=1$.\\
Defining the small parameter $\bar \beta=\beta-1$, the perturbing force per unit mass on body $i$ in an inertial frame is
\be
\ve a_\beta^i=-\frac{2 \bar \beta }{c^2}\sum_{j\neq i} \frac{\mu_j \ve r_{ij}}{r_{ij}^3}\left[\sum_{h\neq i}\frac{\mu_h}{r_{ih}}+\sum_{k\neq j}\frac{\mu_k}{r_{jk}} \right]\,.
\ee
We calculate it in the heliocentric frame $\ve a_ \beta^i-\ve a_\beta^0$. 
The biggest term is $2 \bar \beta /c^2 \mu_0^2  /R_{0i}^3\, \ve u_r^i$ and all others are at least~3 orders of magnitude smaller.\\
The effect due to $\bar \beta$ is essentially a radial force, as for $\delta_{\mu_0}$ and $\zeta$. 
A particular solution is 
\be
x_i  = \bar \beta \dfrac{2 \mu_0^2}{R_{0i}^3 n_i^2 c^2}, \qquad y_i  = - \bar \beta  \dfrac{4  \mu_0^2 }{R_{0i}^3 c^2 n_i }\, t\,;
\label{eq:part_beta}
\ee
and, by Eqs.~(\ref{eq:pert1}), (\ref{eq:pert2})  we obtain the range signature for~$\bar \beta$. 

\subsubsection{Range signature due to Sun's $J_{2\odot}$}
So far, inclinations have been neglected: all planets are assumed to orbit on the ecliptical plane, but in this case we will consider the orbital inclinations with respect to the Sun's equatorial plane. 
This is necessary to avoid a fictitious strong correlation between $J_{2\odot}$ and $\bar \beta$, $\mu_0$ or $\zeta$.\\ 
Inclinations of Mercury and Earth orbits with respect to the Sun's equator are $3.380^\circ$ and $7.155^\circ$ respectively.\\
Unit vectors $\ve u_r$, $\ve u_t$ and $\ve u_z$ have been rewritten to take into account the orbital elements of the planet.\\
The perturbation of Sun's $J_{2\odot}$ on the trajectory of a planet can be obtained by solving this set of equations~\cite{schweighart2001}
\be
\begin{split}
\ddot x-2 (n c) \dot y-(5 c^2-2) n^2 x & = -3 \alpha n^2 (3+5 s)\cos (2 n c t+\varphi)\\
\ddot y+2 (n c) \dot x & =-2 \alpha n^2 (3+5s) \sin (2 n c t +\varphi)\\
\ddot z+(3 c^2-2) n^2 z & =-2 \beta n^2 s \sqrt{1+3 s} \sin (n c t +\varphi )
\end{split}
\ee
where
\be
\begin{split}
c &=\sqrt{1+s};   \qquad  s =\frac{3 J_{2\odot}R_\odot ^2}{8 r^2} \left[1+3 \cos (2 I)\right]; \\
\alpha & =\frac{3 J_{2\odot} R_\odot ^2}{8 r (3+ 5 s)} \left[1-\cos (2 I)\right];   \quad \beta  = \frac{3 J_{2\odot} R_\odot ^2}{4 r s \sqrt{1+3s}} \sin (2 I)\,;
\end{split}
\ee
and $I$ is the inclination, $R_\odot$ is the radius of the Sun, $n$ is the mean motion, $r$ is the Sun-planet distance and $\varphi$ is the initial phase.\\
Hill's equations have been modified by increasing the angular velocity of the reference frame from $n$ up to $n\, c$ in order to avoid drifts into the inhomogeneous solutions [see \cite{schweighart2001} for details]. The particular solution is
\be
\label{eq:pert_j2}
\begin{split}
x' &=\alpha \cos (2 n  c t+\varphi)\,,\\
y' &=\alpha \dfrac{1+3 s}{2 (1+s)}\sin (2 n  c t+\varphi)\,,\\
z' &=-\beta \sqrt{1+3 s}\sin (n c t+ \varphi) \,.\\
\end{split}
\ee
Since \eref{eq:pert_j2}  are not linear functions of $J_{2\odot}$, we expand them as MacLaurin series of $J_{2\odot}$ up to the first order (for  Mercury and Earth). \\
By Eqs.~(\ref{eq:pert1}) and (\ref{eq:pert2}) we estimate the perturbation on the range due to $J_{2\odot}$.

\subsubsection{Time sampling}
Due to the visibility windows, as described in \sref{sec:o14}, range and range-rate data contain several gaps.
A gap  occurs approximately at each arc and lasts about 9.3 hrs.\\
To perform a realistic calculus, we evaluate the perturbations at the set of epochs $t_i$ (spanning an interval of 373~d) generated by {\em ORBIT14}. 
A low-frequency sampling ($f_s=10^{-4}$~Hz) is sufficient for our purposes since signals involved have frequencies of the same order of planetary mean motions. 
For the RMS $\hat \sigma_i$  relative to the $i$-th  range data, we adopt   \cite{MG}
\be
 \hat \sigma_i = 15\,\mbox{cm } \sqrt{300 f_s} \approx 2.6\, \mbox{cm}
\ee

\subsubsection{Constraints and covariance matrix calculus}
As explained in \sref{sec:nominal}, there is a subset of $M$ parameters $\ve x =\{x_1,...,x_M \} \in \ve q$ for which information are available by other experiments. 
In our case they are the GPs and $\gamma$. The information about $\gamma$ from SCE affects $\beta$ and $\eta$ thanks to the Nordtvedt equation.\\
We define the {\em apriori} observations as $\ve x^P$ and $\ve C^P$ as the non-diagonal {\em apriori} normal matrix used to represent the information available about 
parameters $\ve x$.  Errors relative to the {\em apriori} constraints being $\sigma_i$ with $i=1 \dots M$.\\
The constraint involving the subset $\ve x$ can be written as  $\ve C^P \ve x= \ve C^P \ve x^P$.\\
Following \cite{MG}, we modify \eref{eq:1} by  including the constraints
\begin{equation}
  Q(\mathbf{q})=\frac{1}{m+M}\left[ \boldsymbol{\xi}^T(\mathbf{q})
\ve W\boldsymbol{\xi}(\mathbf{q})+ (\ve x-\ve x^P)^T \ve C^P (\ve x- \ve x^P)\right] .
\end{equation}
In our case we have
\be
(\ve x-\ve x^P)^T \ve C^P (\ve x- \ve x^P)=\frac{(\eta-4 \bar \beta)^2}{\sigma_N^2}+\sum_i  \frac{(\mu_i- \mu_i^P)^2}{\sigma_i^2}
\ee
 where $\sigma_N=2.0 \times 10^{-6}$ is the expected RMS of $\gamma$ after SCE, while the summation is extended to all GPs and $\sigma_i$ are the corresponding errors. \\
By means of well known formulas, we obtain the normal matrix $\mathbf C$ and its inverse, the covariance matrix $\mathbf \Gamma$. Finally, the diagonal elements of $\mathbf \Gamma$ give us the expected RMSs of the parameters.\\
For a large fraction of asteroids GPs are estimated by ground-based measurements of diameters and supposed density values, therefore uncertainties are large.\\
However, at the epoch of BC mission, a certain number of GPs will become more precise thanks to GAIA mission by precise measurements of the perturbations on Mars orbit and asteroid-asteroid close approaches.\\
By DAWN measurements the error of 4~Vesta's GP is now $1.2 \times 10^{-5}$~[km$^3$/s$^2$] \cite{konopliv2014}, while the GP of 1~Ceres has been recently estimated with an error of $8.0\times 10^{-4}$~[km$^3$/s$^2$] \cite{park2016}.\\
The current  error $\sigma[\mu_5]=2.7$~[km$^3$/s$^2$] of the GP of the Jupiter system could be improved by JUNO mission measurements up to $\sigma[\mu_5]=0.53$ ($X$-band) or 0.20~[km$^3$/s$^2$] ($K_a$-band)   \cite{serraphd}.\\
In \cite{mouret2009} a list of asteroids with their expected relative errors on GPs after GAIA mission, is reported, 62 of them belong to the sample of 343 asteroids we considered.
We perform tests with both the ``current'' and the ``expected'' values (reported in \tref{tab_curexp}).
\begin{table}[!htbp] 
\begin{ruledtabular}
\begin{tabular}{l d d cc}
\textrm{Body}           & \multicolumn{2}{l}{\textrm{GPs errors  [km$^3$/s$^2$]}}               &        \textrm{  refs}\\
                        &  \textrm{(curr.)}                        & \textrm{(exp.)      }                         &\\ 
\hline
Venus              & 0.0063                    & 0.0063                                      &\cite{luzum2011}\\
Mars               &0.00028                      &0.00028                                   &\cite{luzum2011}\\
Jupiter            & 2.7                              & 0.5                                         & \cite{luzum2011,serraphd}\\
Saturn             & 1.1                              & 1.1                                        &   \cite{luzum2011} \\
Uranus           & 5.0                              & 5.0                                          &  \cite{luzum2011}  \\
Neptune         & 10.5                             & 10.5                                       &   \cite{luzum2011} \\
Pluto              & 2.1                                & 2.1                                      & \cite{stern2015}\\
Eris                 & 13.1                             & 13.1                                     &\cite{luzum2011}\\
1 Ceres          & 0.0008                       & 0.0008                                      & \cite{park2016}\\
2 Pallas          & 0.28                            & 0.17                                        & \cite{luzum2011,mouret2009} \\
3 Juno             & 0.11                            & 0.037                                     & \cite{luzum2011,mouret2009} \\
4 Vesta            & 0.000012                      &  0.000012                                & \cite{konopliv2014} \\
10 Hygiea       &0.48                             & 0.043                                          &\cite{mouret2009}\\
704 Interamnia&0.47                             & 0.11                                          &\cite{mouret2009}\\
\dots               &\dots                             &\dots                                        &\dots    \\
\end{tabular}
\end{ruledtabular}
\caption{\footnotesize Current and ``expected'' uncertainties for  GPs of planets/asteroids. 
For Jupiter, the improvement could be reached by JUNO mission data, for the others in the list, by GAIA.}
\label{tab_curexp}
\end{table}

\subsubsection{Results}
We made a preliminary test to check the results of the analytical model with those of the nominal experiment described in  \sref{sec:nominal}. Therefore, 
we perform the covariance analysis for the following set of parameters only: barycentric initial conditions of Mercury and Earth plus $\bar \beta$, $\eta$, $\delta_{\mu_0}$, $J_{2\odot}$ and $\zeta$.\\
Results are reported in col.~(1) of  \tref{tab_results}, by comparing with the RMSs of \tref{tab_relnom} we find a very good agreement. 
In particular, the outcome for $\eta$ from the analytical model is $\sigma[\eta]=1.58\times 10^{-6}$, to be compared with $1.93\times 10^{-6}$ of {\em ORBIT14}.\\
After the validation of the analytical model, we did two experiments with an extended set of parameters.\\
In both cases we added to the list the GPs of planets from Jupiter to Neptune plus Pluto, Eris and the whole sample of asteroids (343) involved in the dynamics of {\em ORBIT14}.
The total number of parameters was 362. The Nordtvedt apriori has been included in both experiments.\\
In the first experiment, we adopt for the aprioris the {\em current} errors of GPs, while in the other we use the {\em expected} ones, according to values reported in \tref{tab_curexp}. 
Results are reported in columns~(2) and (3) of \tref{tab_results}.\\
The resulting RMS of the parameter $\eta$ is about 1~order of magnitude with respect to the ideal case.\\
The difference between the two experiments (current and expected) is small: the aprioris with the expected values improve the RMS of $\eta$ by only a factor 1.4 leading to $\sigma[\eta]=3.1 \times 10^{-5}$.


\section{Conclusions}\label{sec:concl}
In this work we described in details a semi-analytical model for the Earth-Mercury ranging for the BepiColombo mission.
Our purpose was the  estimation of  the RMS of the Nordtvedt parameter $\eta$ by a global covariance analysis.
The frequency of signal due to $\eta$ in the Earth-Mercury range is of the same order of planetary mean motions and 
  the parameters that could be in principle correlated with $\eta$ are the initial conditions of Earth and Mercury, the other PN parameters, and the masses of 
  planets and asteroids. We included  them in the list of parameters by calculating their signals on the range.\\
In order to check the analytical model we performed a preliminary covariance analysis, involving 13 parameters, to be compared with the numerical global-fit 
obtained by {\em ORBIT14} in the "ideal" case  of exact knowledge of the masses of planets and asteroids.
We  found that the RMSs given by our model were in good agreement with those estimated by  {\em ORBIT14}.\\
Afterwards, we included to the parameters list the masses of planets and the 343 more massive asteroids   (the total number of parameters was 362).
The RMSs of the masses have been constrained to their current (or expected at the epoch of the mission) values and we found 
$\sigma[\eta]=4.37 \times 10^{-5}$ and $\sigma[\eta]=3.13 \times 10^{-5}$, respectively.
Therefore, the uncertainties of the masses of solar system bodies degradate the precision of the estimation of $\eta$  of about 1~order of magnitude.
However, since the current RMS of $\eta$, from LLR measurements, is  $\sigma [\eta]=4.4 \times 10^{-4}$, we conclude that the BepiColombo relativity 
experiment can improve  the current precision on $\eta$ by a factor 10.

\begin{table}[h!]
\begin{tabular}{l l l l l }
\hline
\hline
\multicolumn{5}{c}{Relativity experiment }\\
\multicolumn{5}{c}{(integration time: 373~d)}\\[6pt]
 parameter                                &                  units                                       &              (1)                                                        &                 (2)                                                &                   (3)                               \\  [4pt]
  \hline
 $X_{0,m}$                             &                  [cm]                                        &                   0.29                                                &                 2.51$\times10^3$                        &                 2.49$\times10^3$           \\ 
 $Y_{0,m}$                             &                  [cm]                                        &                   0.88                                               &                   1.19$\times10^4$                        &                 1.18$\times10^4$             \\                                                        
 $Z_{0,m}$                             &                  [cm]                                       &                   4.62                                                &                 5.38                                              &                  5.15	                                \\                                                         
 $\dot X_{0,m}$                        &                  [cm s$^{-1}$]                       &                   3.91$\times10^{-7}$                     &                 2.38$\times 10^{-3 }$                     &                  2.36 $\times 10^{-3 }$         \\                              
 $\dot Y_{0,m}$                        &                  [cm s$^{-1}$]                        &                   2.93$\times10^{-7}$                     &                 1.70$\times 10^{-3 }$                     &                  1.68 $\times 10^{-3 }$         \\                         
 $\dot Z_{0, e}$                        &                  [cm s$^{-1}$]                         &                   3.90$\times10^{-6}$                     &                 4.76$\times 10^{-6 }$                     &                  4.72 $\times 10^{-6 }$         \\                            
 $\dot X_{0, e}$                         &                  [cm s$^{-1}$]                       &                   1.04$\times10^{-7}$                     &                 1.79$\times 10^{-3 }$                     &                  1.77 $\times 10^{-3 }$         \\                          
 $\dot Y_{0, e}$                         &                  [cm s$^{-1}$]                         &                   1.18$\times10^{-7}$                     &                 9.47$\times 10^{-5 }$                     &                 9.41$\times 10^{-5 }$         \\                    
 $\beta$                                       &                  -                                            &                   6.38$\times10^{-7}$                     &                 1.09$\times 10^{-5 }$                     &                  7.81 $\times 10^{-6 }$      \\                       
 $\eta$                                        &                  -                                            &                   1.58$\times10^{-6}$                     &                 4.37$\times 10^{-5 }$                     &                  3.13 $\times 10^{-5 }$       \\              
 $\delta_{\mu_0}$                                      &                  [cm$^3$s$^{-2}$]                &                   9.19$\times10^{12}$                     &                 7.69$\times 10^{13 }$                     &                 5.50 $\times 10^{13 }$       \\                      
 $J_{2\odot}$                                    &                  -                                       &                   3.80$\times10^{-10}$                    &                 8.49$\times 10^{-10}$                     &                 8.03 $\times 10^{-10}$         \\                       
 $\zeta$                                          &                  [yr$^{-1}$]                          &                   1.22$\times10^{-14}$                    &                 1.93$\times 10^{-14}$                     &                 1.78 $\times 10^{-14}$         \\ 
...                                                   &                                                              &                                                                         &                         ...                                 &                     ...                             \\                                                                                   
\hline
\hline
\end{tabular}
\caption{RMSs of the parameters obtained by the analytical global covariance analysis.\\ (1) Metric+SCE experiment assuming no errors on GPs (to be compared with numerical 
results listed in \tref{tab_relnom}). (2) Metric+SCE experiment assuming ''current'' errors on GPs. (3) as (2) but with ''expected'' errors on GPs.}
\label{tab_results}
\end{table}


\begin{acknowledgments}

FDM acknowledges the advice and support  of N.~ Ashby and P.~ Bender (University of Colorado, Boulder) for fruitful interaction on the development of analytical models.\\
The results of the research presented in this work have been performed within the scope of the contract ASI/2007/I/082/06/0 with the Italian Space Agency.

\end{acknowledgments}

\clearpage

\appendix
\onecolumngrid
\section{Matrices}

\be
\label{eq:hatv}
\ve {\hat v}=
\left(
\begin{array}{l l l l l l}
\cos (n t) & \sin (n t) & 1 & 0 & 0 & 0 \\
-2 \sin (n t) & 2 \cos (n t) & -3 n t /2 & 0 & 0 & 0 \\
0 & 0 & 0 & 0 & \cos (n t) & \sin (n t) \\
-n \sin (n t) & n \cos (n t) & 0 & 0 & 0 & 0 \\
-2 n \cos (n t) & -2 n \sin (n t) & -3 n /2 & 0 & 0 & 0 \\
0 & 0 & 0 & 0 & -n  \sin (n t) & n \cos (n t) 
\end{array}
\right)
 \ee

\be
\label{eq:calV}
{\cal V}=\ve {\hat v}\,  \ve {\hat v}_0^{-1} =
\left(
\begin{array}{l l l l l l}
 4 -3 \cos (n t)  &  0  &  0  &  \sin (n t)/n  &  2 \left[1- \cos (n t)\right]/n  &  0  \\
 6 \left[\sin (n t)-n t\right]  &  1  &  0  &  2\left[\cos (n t)-1\right]/n  &  -3 t+ 4 \sin (n t)/n  &  0  \\
 0  &  0  &  \cos (n t)  &  0  &  0  &  \sin (n t)/n  \\
 3 n \sin (n t)  &  0  &  0  &  \cos (n t)  &  2 \sin (n t)  &  0  \\
 6 n \left[\cos (n t)-1\right]  &  0  &  0  &  -2 \sin (n t)  &  -3+4  \cos (n t) &  0  \\
 0  &  0  &  -n \sin (n t)  &  0  &  0  &  \cos (n t)  \\
 \end{array}
\right)
 \ee

\be
\label{eq:R}
{\cal R}=
\left(
\begin{array}{l l l l l l}
\cos (n t) & \sin (n t) & 0 & 0 & 0 & 0 \\
-\sin (n t) &  \cos (n t) & 0 & 0 & 0 & 0 \\
0 & 0 & 1 & 0 &0 & 0  \\
-n \sin (n t) & n \cos (n t) & 0 & \cos (n t) & \sin (n t) & 0 \\
- n \cos (n t) & - n \sin (n t) & 0 & -\sin (n t) & \cos (n t) & 0 \\
0 & 0 & 0 & 0 & 0 & \\
\end{array}
\right)
 \ee

\twocolumngrid

\section{Additional signals due to barycentric initial conditions}
The extra terms to  be added to \eref{eq:pert2}, if initial conditions are barycentric, are
$g^k-g^i$ where

\be
\label{eq:additional}
 g^i= -{\cal R}^{-1} {\cal V} {\cal R}_0
\begin{cases}
\Omega_0 \ve R_0 & \quad q_m=\eta; \\
(R_{0l} \ve s_l+\ve R_0)/\mu_T   & \quad         q_m=      \mu_l \neq \mu_0    ; \\
\ve R_0/\mu_T &     \quad     q_m=  \mu_l = \mu_0    ; \\
\end{cases}
\ee
where  $\mu_T$ is the total mass (Sun, planets and asteroids) and $\cal R, {\cal V}$, ${\ve s}_l$ are calculated for body~$i$.

\section{Complete SEP violation perturbing term}
Complete radial and along-track perturbations on planet $i$ due to SEP violation
\be
\label{eta_exact}
\begin{split}
R_\eta^i &= - \dfrac{\mu_i \Omega_0+\mu_0 \Omega_i}{R_{0i}^2}+\\
 &  +\displaystyle \sum_{j\neq i \neq 0} \mu_j \left(  \Omega_i \dfrac {R_{0j}\cos \Phi_{ij}-R_{0i}}{R_{ij}^3}- \Omega_0 \dfrac{\cos \Phi_{ij}}{R_{0j}^2}\right)\\
T_\eta^i &=    \displaystyle \sum_{j\neq i \neq 0} \mu_j  \sin \Phi_{ij} \left(\Omega_i \dfrac {R_{0j}}{R_{ij}^3}- \Omega_0 \dfrac{1}{R_{0j}^2}\right)
\end{split}
\ee
\section{Coefficients for planetary perturbations}
Series expansion for $1/R_{ij}^3$, here $R_{0i}<R_{0j}$

\be
\label{eq:r3}
\begin{split}
 \frac{1}{R_{ij}^3} &= \left(\frac{1}{R_{0j}^3}+\frac{9}{4}\frac{R_{0i}^2}{R_{0j}^5}+\dots \right)+ \\
 &+ \left(3\,\frac{R_{0i}}{R_{0j}^4}+\frac{45}{8}\frac{R_{0i}^3}{R_{0j}^6}+\dots\right)\cos \Phi_{ij} +\\
 & + \left(\frac{15}{4}\,\frac{R_{0i}^2}{R_{0j}^5}+\frac{105}{16}\frac{R_{0i}^4}{R_{0j}^7}+\dots\right) \cos 2 \Phi_{ij}+\dots
\end{split}
\ee

Radial ($a_{j,l}$) and along-track ($b_{j,l}$) coefficients of the perturbation on body $i$ due to body $j$ [see \eref{eq:plan}]
\be
\label{eq:ajbj}
\begin{split}
 a_{j,0} & = \dfrac{1}{2}\dfrac{R_{0i}}{ R_{0j}^3}+\dfrac{9}{16}\dfrac{R_{0i}^3}{R_{0j}^5}+\dfrac{75}{128}\dfrac{R_{0i}^5}{R_{0j}^7}\dots ;\\
 a_{j,1}  &=  \dfrac{9}{8}\dfrac{ R_{0i}^2}{ R_{0j}^4}+ \dfrac{75}{64}\dfrac{R_{0i}^4}{ R_{0j}^6} + \dfrac{305}{512}\dfrac{ R_{0i}^6}{ R_{0j}^8}+\cdots;\\ 
 a_{j,2}  &= \dfrac{3}{2}\dfrac{R_{0i}}{R_{0j}^3}+\dfrac{5}{4}\dfrac{R_{0i}^3}{R_{0j}^5}+\dfrac{315}{256}\dfrac{R_{0i}^5}{R_{0j}^4} +\cdots;\\
 b_{j,1}  &= \dfrac{3}{8} \dfrac{ R_{0i}^2}{ R_{0j}^4}+ \dfrac{15}{64}\dfrac{ R_{0i}^4}{ R_{0j}^6}+ \dfrac{95}{512}\dfrac{ R_{0i}^6}{ R_{0j}^8}+\cdots; \\
 b_{j,2}  &= \dfrac{3}{2}\dfrac{ R_{0i}}{ R_{0j}^3}+\dfrac{5}{8}\dfrac{ R_{0i}^3}{ R_{0j}^5}+\dfrac{105}{256}\dfrac{ R_{0i}^5}{ R_{0j}^7}+\cdots; \\
\end{split}
\ee

\onecolumngrid
\section{Amplitudes and frequencies for  SEP violation and planetary signatures}

\begin{table*}[h!]
\begin{ruledtabular}
 \begin{tabular}{l rr rrrrrrrrrr} 
   & \multicolumn{2}{c}{Direct terms (${\cal D}_l$)} &  \multicolumn{10}{c}{Indirect terms (${\cal I}_{pqr}$)}\\[8pt]
                                &                                                &                           & \multicolumn{2}{c}{Mars}  & \multicolumn{2}{c}{Jupiter} & \multicolumn{2}{c}{Saturn} & \multicolumn{2}{c}{Uranus} & \multicolumn{2}{c}{Neptune}\\[8pt]
    frequency                      & period         & ampl \hspace*{0.5cm}& period  & ampl  & period  & ampl  & period  & ampl  & period  & ampl  & period  & ampl  \\
                                          &           [d]       &    [m] \hspace*{0.5cm}&    [d] & [m] & [d] & [m] & [d] & [m] & [d] & [m] & [d] & [m] \\[4pt]
\hline
     0                                       & $\infty$   & -22.48  \hspace*{0.5cm}          &     -     &     -     &     -     &     -     &     -     &     -     &     -     &        -     &   - & -\\
$n_1-n_3$                              & 115.9     & 1.49 \hspace*{0.5cm}              &     -     &     -     &     -     &     -     &     -     &     -     &     -     &        -     &   - &- \\
$n_j-n_3$                               &              &                                                  &  747.3&  0.36   &  398.8  &  232.81  &  378.09  &  47.34  &  369.66  &  4.92  &  367.5  &  4.59  \\
$n_1-n_j$                               &              &                                                  & 100.3 &  -0.30  &  89.8  &  -137.31  &  88.7  &  -27.27  &  88.2  &  -2.81  &  88.1  &  -2.61  \\
$n_1+n_j-2 n_3$                    &            &                                                    &   137.1 & 0.28  &  163.3  &  118.88  &  167.1  &  23.42  &  168.8  &  2.40  &  169.2  &  2.23  \\
$2 n_1-2 n_3$                        & 57.9      & -0.56 \hspace*{0.5cm}              &     -     &     -     &     -     &     -     &     -     &     -     &     -     &     -     &     -     &   -\\
$2 n_1+n_j-3 n_3$                  &            &                                                   &  62.8  & 0.06  &  67.8  &  29.04  &  68.4  &  5.76  &  68.7  &  0.59  &  68.8  &  0.55  \\
$2 n_1-n_j-n_3$                     &            &                                                    &  53.8  &  -0.02  &  50.6  &  20.36  &  50.2  &  4.74  &  50.1  &  0.52  &  50.05  &  0.49  \\
$3 n_1-3 n_3$                        & 38.6      & -0.19   \hspace*{0.5cm}            &     -     &     -     &     -     &     -     &     -     &     -     &     -     &     -     &     -     &   -\\
$3 n_1+n_j-4 n_3$                 &              &                                                  &  40.7  &  0.01  &  42.8  &  4.47  &  43.0  &  0.87  &  43.1  &  0.09  &  43.2  &  0.08  \\
$3 n_1-n_j-2 n_3$                  &              &                                                  &  36.7  &  -0.01  &  35.2  &  -2.56  &  35.0  &  -0.40  &  35.0  &  -0.04  &  35.0  &  -0.03  \\
$4 n_1-4 n_3$                        & 29.0      & -0.12 \hspace*{0.5cm}              &     -     &     -     &     -     &     -     &     -     &     -     &     -     &     -     &     -     &   -\\
$4 n_1-n_j-3 n_3$                  &               &                                                 &  27.9  &  -  &  27.0  &  2.35  &  26.9  &  0.51  &  26.9  &  0.05  &  26.9  &  0.05  \\
$2 n_j-2 n_3$                          &                &                                               &  373.7  &  -  &  199.4  &  -  &  189.0  &  -  &  184.8  &  -  &  183.7  &  -  \\
$n_1-2 n_j+n_3$                     &                &                                                &  88.4  &  -  &  73.3  &  -  &  71.8  &  -  &  71.2  &  -  &  71.1  &  -  \\
$n_1+2 n_j-3 n_3$                  &                 &                                                &  168.0  &  -  &  276.6  &  -  &  299.4  &  -  &  310.6  &  -  &  313.7  &  -  \\
$5 n_1-5 n_3$                        & 23.2      & 0.06 \hspace*{0.5cm}               &     -     &     -     &     -     &     -     &     -     &     -     &     -     &     -     &     -     & - \\
...                                           & ...           & ...     \hspace*{0.5cm}               & ... & ... & ... & ... & ... & ... & ... & ... & ... & ...\\
\end{tabular}
\end{ruledtabular}
 \caption{SEP violation range signature: coefficients relative to the direct and indirect parts (for $\eta=1$).  Only terms bigger than 1 cm are reported. See \eref{eq:eta}.} 
 \label{tab:eta}
\end{table*}

\begin{table*}
\begin{ruledtabular}
 \begin{tabular}{l r r r r r r r r r r} 
  \multicolumn{11}{c}{Coefficients ${\cal J}$ of Earth-Mercury range perturbation}\\
 & \multicolumn{2}{c}{Mars}  & \multicolumn{2}{c}{Jupiter} & \multicolumn{2}{c}{Saturn} & \multicolumn{2}{c}{Uranus} & \multicolumn{2}{c}{Neptune} \\
frequency  & period  & ampl  & period  & ampl  & period  & ampl  & period  & ampl  & period  & ampl \\
 & [d] & [m] & [d] & [m] & [d] & [m] & [d] & [m] & [d] & [m] \\
\hline
$0$ & $\infty $ &       -1178.5       & $\infty $ &       -162393.6       & $\infty $ &       -7941.0       & $\infty $ &       -149.2       & $\infty $ &       -45.7       \\
$5 n_j-5 n_3$ &       149.5       &       -759.0       &       79.8       &       -345.1       &       75.6       &       -2.4       &       73.9       &  -  &       73.5       &  -  \\
$4 n_j-4 n_3$ &       186.8       &       -4089.4       &       99.7       &       -5517.9       &       94.5       &       -70.7       &       92.4       &       -0.3       &       91.9       &  -  \\
$3 n_j-3 n_3$ &       249.1       &       -30352.7       &       132.9       &       -92266.9       &       126.0       &       -2127.8       &       123.2       &       -18.7       &       122.5       &       -3.6       \\
$2 n_j-2 n_3$ &       373.7       &       1405238.7       &       199.4       &       -1303037.9       &       189.0       &       -52886.5       &       184.8       &       -921.9       &       183.7       &       -277.2       \\
$n_j-n_3$ &       747.3       &       30357.6       &       398.8       &       2143106.1       &       378.1       &       134196.8       &       369.7       &       3498.7       &       367.5       &       1334.9       \\
$n_1+5 n_j-6 n_3$ &             515.6       &       -184.3       &       255.9       &       -68.9       &       217.6       &       -0.5       &       204.2       &  -  &       201.0       &  -  \\
$n_1+4 n_j-5 n_3$ &       305.1       &       -1098.4       &       714.2       &       -1154.8       &       512.9       &       -14.5       &       456.4       &  -  &       443.5       &  -  \\
$n_1+3 n_j-4 n_3$ &       216.7       &       -9127.3       &       903.1       &       -20781.5       &       1438.5       &       -469.6       &       1944.8       &       -4.10       &       2144.8       &       -0.8       \\
$n_1+2 n_j-3 n_3$ &       168.0       &       535946.3       &       276.6       &       -343492.5       &       299.4       &       -13562.0       &       310.6       &       -233.8       &       313.7       &       -70.1       \\
$n_1+n_j-2 n_3$ &       137.1       &       20384.6       &       163.3       &       892047.9       &       167.1       &       53440.7       &       168.8       &       1367.1       &       169.2       &       519.0       \\
$n_1-n_3$ &       115.9       &       88.2       &       115.9       &       9656.0       &       115.9       &       471.7       &       115.9       &       8.9       &       115.9       &       2.7       \\
$n_1-n_j$ &       100.3       &       -21612.9       &       89.8       &       -982233.0       &       88.7       &       -59100.9       &       88.2       &       -1514.8       &       88.10       &       -575.4       \\
$n_1-2 n_j+n_3$ &       88.4       &       -588309.0       &       73.3       &       422347.3       &       71.8       &       16944.1       &       71.2       &       294.1       &       71.1       &       88.3       \\
$n_1-3 n_j+2 n_3$ &       79.1       &       10278.3       &       61.9       &       24391.7       &       60.4       &       553.2       &       59.7       &       4.8       &       59.5       &       0.9       \\
$n_1-4 n_j+3 n_3$ &       71.5       &       1252.6       &       53.6       &       1363.6       &       52.1       &       17.2       &       51.4       &  -  &       51.2       &  -  \\
$n_1-5 n_j+4 n_3$ &       65.3       &       212.9       &       47.2       &       81.9       &       45.8       &       0.6       &       45.1       &  -  &       45.0       &  -  \\
$2 n_1+5 n_j-7 n_3$ &       94.6       &       -53.9       &       211.8       &       -21.6       &       247.8       &       -0.2       &       267.8       &  -  &       273.7       &  -  \\
$2 n_1+4 n_j-6 n_3$ &       84.0       &       -311.4       &       138.3       &       -355.6       &       149.7       &       -4.5       &       155.3       &  -  &       156.9       &  -  \\
$2 n_1+3 n_j-5 n_3$ &       75.5       &       -2503.8       &       102.7       &       -6238.2       &       107.2       &       -142.0       &       109.4       &       -1.2       &       109.9       &       -0.2       \\
$2 n_1+2 n_j-4 n_3$ &       68.6       &       138269.1       &       81.7       &       -98387.3       &       83.5       &       -3921.1       &       84.4       &       -67.9       &       84.6       &       -20.4       \\
$2 n_1+n_j-3 n_3$ &       62.8       &       4726.0       &       67.8       &       225809.9       &       68.4       &       13663.2       &       68.7       &       351.0       &       68.8       &       133.4       \\
$2 n_1-2 n_3$ &       57.9       &       -38.5       &       57.9       &       -5722.2       &       57.9       &       -279.9       &       57.9       &       -5.3       &       57.9       &       -1.6       \\
$2 n_1-n_j-n_3$ &       53.8       &       -3189.6       &       50.6       &       -102840.3       &       50.2       &       -5908.8       &       50.1       &       -148.3       &       50.0       &       -56.0       \\
$2 n_1-2 n_j$ &       50.2       &       -83674.5       &       44.9       &       42429.8       &       44.3       &       1617.2       &       44.1       &       27.4       &       44.0       &       8.2       \\
$2 n_1-3 n_j+n_3$ &       47.0       &       1364.9       &       40.4       &       3385.9       &       39.7       &       78.1       &       39.4       &       0.7       &       39.3       &       0.1       \\
$2 n_1-4 n_j+2 n_3$ &       44.2       &       152.8       &       36.6       &       143.9       &       35.9       &       1.8       &       35.6       &  -  &       35.5       &  -  \\
$2 n_1-5 n_j+3 n_3$ &       41.8       &       24.4       &       33.6       &       8.2       &       32.8       &  -  &       32.5       &  -  &       32.4       &  -  \\
$3 n_1+5 n_j-8 n_3$ &       52.1       &       -2.8       &       74.9       &       -0.4       &       79.0       &  -  &       80.9       &  -  &       81.4       &  -  \\
$3 n_1+4 n_j-7 n_3$ &       48.7       &       -21.1       &       63.0       &       -9.7       &       65.3       &       -0.1       &       66.4       &  -  &       66.6       &  -  \\
$3 n_1+3 n_j-6 n_3$ &       45.7       &       -212.4       &       54.4       &       -246.6       &       55.7       &       -5.1       &       56.3       &  -  &       56.4       &  -  \\
$3 n_1+2 n_j-5 n_3$ &       43.1       &       16327.3       &       47.9       &       -6393.2       &       48.5       &       -238.0       &       48.8       &       -4.0       &       48.9       &       -1.2       \\
$3 n_1+n_j-4 n_3$ &       40.7       &       857.4       &       42.8       &       29221.8       &       43.0       &       1691.4       &       43.1       &       42.6       &       43.2       &       16.1       \\
$3 n_1-3 n_3$ &       38.6       &       29.0       &       38.6       &       3864.6       &       38.6       &       188.9       &       38.6       &       3.6       &       38.6       &       1.1       \\
$3 n_1-n_j-2 n_3$ &       36.7       &       -1480.3       &       35.2       &       -70614.5       &       35.0       &       -4273.7       &       35.0       &       -109.8       &       35.0       &       -41.7       \\
$3 n_1-2 n_j-n_3$ &       35.0       &       -48191.0       &       32.4       &       30888.6       &       32.1       &       1205.5       &       31.9       &       20.7       &       31.9       &       6.2       \\
$3 n_1-3 n_j$ &       33.4       &       890.3       &       29.9       &       2187.9       &       29.6       &       49.5       &       29.4       &       0.4       &       29.4       &  -  \\
$3 n_1-4 n_j+n_3$ &       32.0       &       116.6       &       27.8       &       151.3       &       27.4       &       1.9       &       27.2       &  -  &       27.2       &  -  \\
$3 n_1-5 n_j+2 n_3$ &       30.7       &       20.0       &       26.0       &       8.3       &       25.6       &  -  &       25.4       &  -  &       25.3       &  -  \\
$4 n_1-n_j-3 n_3$ &       27.9       &       20.7       &       27.0       &       2307.8       &       26.9       &       147.7       &       26.9       &       3.9       &       26.9       &       1.5       \\
$4 n_1-2 n_j-2 n_3$ &       26.9       &       -29.7       &       25.3       &       -1437.2       &       25.1       &       -66.5       &       25.0       &       -1.2       &       25.0       &       -0.4       \\
$4 n_1-3 n_j-n_3$ &       26.0       &       -5.6       &       23.8       &       -83.9       &       23.6       &       -2.1       &       23.5       &  -  &       23.4       &  -  \\
$4 n_1-4 n_j$ &       25.1       &       -0.6       &       22.4       &       -3.4       &       22.2       &  -  &       22.1       &  -  &       22.0       &  -  \\
$4 n_1-5 n_j+n_3$ &       24.3       &       0.2       &       21.3       &       0.4       &       20.9       &  -  &       20.8       &  -  &       20.8       &  -  \\
$5 n_1-2 n_j-3 n_3$ &       21.8       &       -5.8       &       20.8       &       -264.1       &       20.6       &       -12.1       &       20.6       &       -0.2       &       20.6       &  -  \\
$5 n_1-3 n_j-2 n_3$ &       21.2       &       -1.8       &       19.7       &       -29.1       &       19.6       &       -0.7       &       19.5       &  -  &       19.5       &  -  \\
$5 n_1-4 n_j-n_3$ &       20.6       &       -0.3       &       18.8       &       -1.4       &       18.6       &  -  &       18.5       &  -  &       18.5       &  -  \\
$5 n_1-5 n_j$ &       20.1       &  -  &       18.0       &  -  &       17.7       &  -  &       17.6       &  -  &       17.6       &  -  \\
$6 n_1-3 n_j-3 n_3$ &       17.9       &       -0.3       &       16.9       &       -4.1       &       16.7       &       -0.1       &       16.7       &  -  &       16.7       &  -  \\
$6 n_1-4 n_j-2 n_3$ &       17.5       &       -0.1       &       16.2       &       -0.6       &       16.0       &  -  &       16.0       &  -  &       16.0       &  -  \\
$6 n_1-5 n_j-n_3$ &       17.1       &  -  &       15.5       &  -  &       15.4       &  -  &       15.3       &  -  &       15.3       &  -  \\
$7 n_1-4 n_j-3 n_3$ &       15.2       &  -  &       14.2       &  -  &       14.1       &  -  &       14.0       &  -  &       14.0       &  -  \\
$7 n_1-5 n_j-2 n_3$ &       14.9       &  -  &       13.7       &  -  &       13.6       &  -  &       13.5       &  -  &       13.5       &  -  \\
$8 n_1-5 n_j-3 n_3$ &       13.2       &  -  &       12.3       &  -  &       12.2       &  -  &       12.1       &  -  &       12.1       &  -  \\
 .. & ... & ... & ... & ... & ... & ... & ... & ... & ... & ...\\
\end{tabular}
 \end{ruledtabular}
\caption{\footnotesize Earth-Mercury range perturbation coefficients due to planetary effects. Only terms bigger than 10 cm are reported.}
\label{tab:pla}
\end{table*}

\twocolumngrid
\clearpage

\bibliography{BCrelativity_refs}

\end{document}